\documentclass[twocolumn]{aastex631}

\usepackage{sidecap}
\usepackage{capt-of}
\usepackage{xcolor}
\usepackage{hhline,soul}
\usepackage{booktabs,tabulary}
\usepackage[flushleft]{threeparttable}
\usepackage{wrapfig}
\usepackage{hepparticles}
\usepackage{breqn}

\newcommand{\TXS}{TXS~0506$+$056}
\newcommand{\origin}{\textsc{Paper~I}}
\newcommand{\originII}{\textsc{Paper~II}}

\shorttitle{Extragalactic neutrino factories}
\shortauthors{Buson et al.}

\graphicspath{{./}{./}}

\begin{document}

\title{Extragalactic neutrino factories}

\author[0000-0002-3308-324X]{Sara Buson}
\affiliation{Fakult\"{a}t f\"ur Physik und Astronomie, Julius-Maximilians-Universit\"at W\"urzburg,\\ Emil-Fischer-St. 31, D-97074, W\"urzburg, Germany}
\email{sara.buson@uni-wuerzburg.com}

\author[000 0-0002-8186-3793]{Andrea Tramacere}
\affiliation{Department of Astronomy, University of Geneva, Ch. d'\`Ecogia 16, Versoix, 1290, Switzerland}

\author[0000-0003-4519-4796]{Lenz Oswald}
\affiliation{Fakult\"{a}t f\"ur Physik und Astronomie, Julius-Maximilians-Universit\"at W\"urzburg,\\ Emil-Fischer-St. 31, D-97074, W\"urzburg, Germany}

\author[0000-0003-4704-680X]{Eleonora Barbano}
\affiliation{Fakult\"{a}t f\"ur Physik und Astronomie, Julius-Maximilians-Universit\"at W\"urzburg,\\ Emil-Fischer-St. 31, D-97074, W\"urzburg, Germany}

\author{Ga\"{e}tan Fichet de Clairfontaine}
\affiliation{Fakult\"{a}t f\"ur Physik und Astronomie, Julius-Maximilians-Universit\"at W\"urzburg,\\ Emil-Fischer-St. 31, D-97074, W\"urzburg, Germany}

\author[0000-0003-2497-6836]{Leonard Pfeiffer}
\affiliation{Fakult\"{a}t f\"ur Physik und Astronomie, Julius-Maximilians-Universit\"at W\"urzburg,\\ Emil-Fischer-St. 31, D-97074, W\"urzburg, Germany}

\author[0000-0002-2515-1353]{Alessandra Azzollini}
\affiliation{Fakult\"{a}t f\"ur Physik und Astronomie, Julius-Maximilians-Universit\"at W\"urzburg,\\ Emil-Fischer-St. 31, D-97074, W\"urzburg, Germany}

\author{Vardan Baghmanyan}
\affiliation{Fakult\"{a}t f\"ur Physik und Astronomie, Julius-Maximilians-Universit\"at W\"urzburg,\\ Emil-Fischer-St. 31, D-97074, W\"urzburg, Germany}

\author[0000-0002-6584-1703]{Marco Ajello}
\affiliation{Department of Physics and Astronomy, Clemson University, Kinard Lab of Physics, Clemson, SC 29634-0978, USA}



\begin{abstract}
Identifying the astrophysical sources responsible for the high-energy cosmic neutrinos has been a longstanding challenge. In a previous work, we report evidence for a spatial correlation between blazars from the 5th Roma-BZCat catalog and neutrino data of the highest detectable energies, i.e. $\gtrsim0.1$~PeV, collected by the IceCube Observatory in the southern celestial hemisphere. The statistical significance 
is found at the level of $2 \times 10^{-6}$ post-trial. 
In this work we test whether a similar correlation exists in the northern hemisphere, were IceCube is mostly sensitive to $\lesssim0.1$~PeV energies. We find a consistent correlation
between blazars and northern neutrino data at the pre-trial p-value of $5.12 \times 10^{-4}$, and a post-trial chance probability of $6.79 \times 10^{-3}$. Combining the post-trial probabilities observed for the southern and northern experiments
yields 
a global post-trial chance probability of $2.59 \times 10^{-7}$ for the genuineness of such correlation. This implies that the spatial correlation is highly unlikely to arise by chance.
Our studies push forward an all-sky subset of 52 objects as highly likely PeVatron extragalactic accelerators.
\end{abstract}



\section{Introduction} \label{sec:intro}
Despite the evidence of a diffuse neutrino flux consistent with an astrophysical origin reported by the IceCube Observatory in the $\gtrsim 100$~TeV to $\sim10$~PeV energy range \citep{icecube2013,IC_north_hard_spectrum:2016}, little is known about the astrophysical objects that are responsible for it. Identifying the astrophysical sources originating the high-energy astrophysical neutrinos represents one of the most compelling questions in multi-messenger astrophysics. Evidence that at least a fraction of the IceCube high-energy ($\gtrsim 0.1$~PeV) neutrinos originates from a peculiar type of active galactic nuclei (AGN), i.e. accreting super-massive black holes at the centers of galaxies that emit non thermal radiation, has been recently provided \citep{Buson:2022,Buson_erratum:2022}. 
Among them, some can launch powerful outflows, and few are capable of accelerating particles to nearly the speed of light in narrow collimated beams, i.e. relativistic jets. The blazar class commonly encompasses those AGN with a relativistic jet closely aligned with Earth and has been widely investigated in previous multi-messenger studies in relation to neutrinos.
Generally, limited multi-wavelength information is available to access the nature of an AGN and, for a given survey, most of the times this is available only for the brightest objects. Moreover, variability may affect the behaviour of a source observed at a given time, making an unambiguous classification not possible. As a result, the classification of AGN  often attempts to categorize the properties observed in a specific waveband (or survey) at the risk of lacking a comprehensive view of the actual physical characteristics of the system \citep[][]{Padovani2017}.
Motivated by these considerations, in our work we search for candidate neutrino counterparts among AGN whose emission exhibits the presence of non thermal radiation over the electromagnetic spectrum, as classified by the 5th Roma-BZCat catalog \citep[5BZCat,][and references therein]{massaro20155th}.

Our previous study  \citep[hereafter \origin,][]{Buson:2022}
focuses on the time-integrated information encoded in the IceCube 7 yr sky map \citep{IceCube7y:2017} and reports the discovery of a statistically significant correlation between $\gtrsim 0.1$~PeV  neutrinos and a sample of 5BZCat blazars located in the southern celestial hemisphere. The study pinpoints a subset of 10 objects, i.e. PeVatron blazars, as highly-likely neutrino emitters.  
This paper, hereafter \originII, represents a complement of \origin. It builds on the findings of \origin\ and expands the investigation to the northern hemisphere using the latest released IceCube $\sim$9-year northern sky map \citep{IC_10ydata_reprocessed:2022}.

The paper is organized as follows.  Section \ref{sec:primer} summarises the hypothesis behind our study, already introduced in Paper I. Section \ref{sec:icecube} presents the IceCube observatory and the neutrino data available publicly at the time of the writing. The neutrino data and the blazar sample used in \originII\ are presented in Section \ref{sec:samples}. Section \ref{sec:cross-corr} addresses the northern hemisphere statistical analysis and discusses the findings in the context of the southern hemisphere results. Section \ref{sec:previous_searches} places the findings in the context of previous all-sky searches for neutrino point sources. Section \ref{sec:pevatron_blazars} discusses in brief the properties of the northern blazars associated with neutrinos. Section \ref{sec:pev_events} provides supportive grounds regarding the astrophysical neutrino energies involved. Section \ref{sec:beyond} discusses the results in a broader context and Section \ref{sec:conclusions} presents the discussion and conclusions.
When referring to the time integration of an analysis (e.g. skymap), the paper refers to the instrument exposure rather than the calendar time span of the observations, i.e. the IceCube livetime, rounded to the integer in years, unless differently specified. When addressing the statistical significance in ``$\sigma$'', we refer to Gaussian standard equivalent (one-sided).

\section{The PeVatron blazars \\ hadronic primer}\label{sec:primer} 
The study embraced in Paper I and Paper II builds on the hypothesis that, if blazars are capable of accelerating cosmic rays, the bulk of the energy-output in form of accompanying neutrino emission resides at energies $\gtrsim1$~PeV. 
This is consistent with the majority of the state-of-the-art blazar theoretical models, i.e. both hadronic and lepto-hadronic models \citep[e.g.][]{Mannheim1993,Mannheim:1995,Atoyan:2003,Stecker:2013,Dermer:2014,Murase_external_fields:2014,Petropoulou:2015,Reimer:2015,Keivani:2018,Cerruti:2019,Rodrigues:2021}.
PeV energies offer a favorable observational window also due to the characteristic of the expected background spectrum, according to which, the blazar spectrum is predicted to cross the steeply falling background one at $\gtrsim 0.1$~PeV \citep[e.g.][]{Mannheim1993}.

Due to the observatory's geographical location and Earth's opacity hampering the detection of high-energy neutrinos ($\gtrsim0.1$~PeV), the astrophysical neutrino energy ranges accessible by the IceCube observatory depend on the celestial declination. While the IceCube southern hemisphere is democratic in energies, i.e. astrophysical neutrinos of all energies can travel unimpeded toward the observatory and be detected, the northern hemisphere is limited to observe astrophysical neutrinos in the TeV/sub-PeV energy range, with the higher-energy neutrinos preferentially detected close to the celestial horizon \citep{IceCube_earth_cross_section:2017,Bustamante:2019}.

In \origin\ we focus on the southern hemisphere, that is sensitive to the highest energies accessible, and observe a statistically significant correlation between IceCube neutrinos and blazars. To test whether the correlation may hold at the lower observed energies, in \originII\ we apply the same analysis of \origin\ to the northern neutrino data. The latter is independent from the one used in \origin\ (see Section \ref{sec:samples}). Testing the hypothesis of blazars as long-term ($\gtrsim$~year-long) neutrino emitters in an independent dataset, offers the opportunity of corroborating the findings of \origin. Besides, performing the same statistical analysis in the northern hemisphere, which concerns mainly muon-proxy energies in the range $\sim 1$~TeV to $\lesssim100 $~TeV, provides an informative estimate of the level of such correlation in a different energy band.

\section{The IceCube Observatory data releases}
\label{sec:icecube}
The most sensitive neutrino detector currently operating is the IceCube observatory, a cubic-kilometer sized Cherenkov detector embedded in the ice at the geographic South Pole. It detects neutrinos above TeV energies by measuring the Cherenkov light produced by charged particles induced by neutrino interactions. The detector is equipped with a total of 5160 optical photomultiplier tubes arranged in 86 cable strings. The final instrument configuration with 86 strings was completed in December 2010 while, prior to that date, data were recorded with partial detector configurations (i.e. 40, 59, 79 strings).

Several works in the literature have previously addressed the search for all-sky, neutrino point-like sources exploiting the time-integrated IceCube data. During the past decade, the IceCube collaboration has reported such searches carried out with the progressive accumulation of more data, different energy ranges, and progressive improvements in the data calibration, reconstruction and statistical analysis techniques, 
e.g. $\sim3$ years analysis \citep{IceCube3y:2013,South_IC_Antares_combined:2016}, 
$\sim4$ years \citep{aartsen2014searches},
$\sim7$ years \citep{IceCube7y:2017,Coenders_PhDT:2016},
$\sim8$ years \citep{IceCube8yNorth:2019}, 
$\sim10$ years \citep[2008-2018,][]{IceCube10y:2020},
$\sim9$ years \citep[2011-2020,][]{IceCube_10y_reprocessed:2022}. 

The muon-track data collected by IceCube across the first  $\sim$~10-yr (2008-2018) of observations have been publicly released in the form of a list of events, along with information regarding the instrument responce functions, such as effective area, reconstructed angular error and energy estimation \citep{IceCube10y_data:2021}.
In principle, using this information one could reproduce searches similar to those carried out by the collaboration.  
Similar releases allowed independent authors to e.g. carry out point-source studies, although such works could not exploit the full information about the instrument because not publicly available \citep[e.g.][]{Hooper:2019,smith:2021,Bei_noradio_corr:2021}.

Moreover, thanks to recent developments by the IceCube collaboration, most of the results based on the publicly released  $\sim$~10-yr event dataset \citep[e.g.,][]{IceCube10y:2020} are nowadays superseded by those that build on the newer event reconstruction and data calibration methods \citep{Bellenghi:2021}. The latter have yet shown their potential, leading to the claim for evidence of neutrino emission from the AGN NGC~1068 \citep{IceCube_10y_reprocessed:2022}.
In this most recent work, \citet{IceCube_10y_reprocessed:2022} presents also an updated analysis of the northern (celestial) sky.

An alternative is to utilise higher-level data products which are released by the IceCube collaboration to the scientific community, when available. Among those concerning all-sky point-searches, two have been released at the time of the writing. These are a 7-y all-sky map \citep{IC_7ydata:2019} and, recently, the aforementioned $\sim$9 years sky map of the northern celestial hemisphere \citep{IceCube_10y_reprocessed:2022}. The latter encompasses timescales similar to the one previously investigated in \citet{IceCube10y:2020}, but employs a reprocessed muon-track dataset based on the full 86-string configuration and improved statistical analysis methods.
 This is expected to bring significant improvements in the analysis. Among other differences, the latest analysis provides improved source localization, flux characterization and thereby discovery potential (by up to 30\% in the case of hard-spectrum signals, $\propto$~E$^{-2}$) over previous works \citep[see e.g.][]{Bellenghi:2021,IceCube_10y_reprocessed:2022,IC_10ydata_reprocessed:2022}. As well as it enables an unbiased estimation of the source parameters.

\subsection{Neutrino sky map data employed in our studies}
In our first investigation (\origin), we employed the IceCube 7-y sky map data \citep{IC_7ydata:2019,IceCube7y:2017}, as it was the only sky map publicly accessible when performing the study. 
We explored the southern hemisphere dataset, which was best suited to test the blazar hypothesis (see Section \ref{sec:primer} and \origin).

Shortly after the publication of \origin, an updated IceCube $\sim9$-y sky map analysis of the northern hemisphere has been released publicly \citep{IC_10ydata_reprocessed:2022}. Among the three northern-sky maps presented in the companion publication \citep{IceCube_10y_reprocessed:2022}, the scan obtained by testing a signal hypothesis with spectral index treated as free parameter
is publicly accessible \citep[][at the time of the writing]{IC_10ydata_reprocessed:2022}.
Compared to the 7-yr northern-sky map, the newest $\sim$9-y northern-sky map is based on better-reconstructed lower-level data and better parametrization of variables in the likelihood framework (see previous section). 
Therefore, to expand our investigation to the northern celestial hemisphere, we employ this improved $\sim$9-y northern-sky map. 

The $\sim9$-y northern sky map ($-3^{\circ} \leq \delta \leq 81^{\circ}$) used in \originII, and $\sim7$ years southern ($ -85^{\circ} < \delta < -5^{\circ}$)  hemisphere sky map used in \origin,  may be regarded as distinct neutrino datasets due to the non-overlapping, complementary sky coverage, differences in the energy range, event-data calibration and reconstruction, background characteristics and analysis techniques applied to the individual events \citep{IceCube7y:2017,Buson:2022,IceCube_10y_reprocessed:2022}.

\section{The Neutrino and Blazar Samples}
\label{sec:samples}

\subsection{Neutrino northern $\sim9$ yr sky map}
The northern hemisphere sky map employed in this study is based on the neutrino data collected by the IceCube observatory between 13 May 2011 and 29 May 2020, for a total exposure time of 3186 days. It uses more than 670\,000 individual events, the largest majority of them are due to background. The northern-sky map is generated by applying a maximum likelihood analysis where the neutrino data are optimized for searches of point-like neutrino sources with declination  $-3^{\circ} \leq \delta \leq 81^{\circ}$.
The overwhelming majority of the events of this dataset are observed at the lower energies, with a median angular resolution of 1.2$^{\circ}$ \citep{IceCube_10y_reprocessed:2022}.

The sky map provides a probability (a local $p$-value), for each direction of the sky at declination, $-3^{\circ} \leq \delta \leq 81^{\circ}$, in a grid of $\sim$($0.2^{\circ} \times 0.2^{\circ}$) pixels. The local p-values are derived by applying the maximum likelihood technique to the event data where the model assumes for the putative point-source a power-law spectrum 
in a given direction of the sky. 
They indicate the level of clustering in the neutrino data, i.e. a measure of the significance of neutrino events being uniformly distributed. The p-values obtained in such manner yet incorporate the declination-dependence effect of the IceCube data \citep{Coenders_PhDT:2016}. They should not be confused with the p-values derived from our statistical analysis (see Section \ref{sec:cross-corr}).  We define the negative logarithm of the provided local p-value as $L = - $log($p$-value). The local $L$-values range from 0 to 6.75; larger values of $L$ imply a larger inconsistency with background expectations and, hence, a higher probability that astrophysical sources are responsible for the spatial clustering of the neutrino events.
%

\subsection{Blazar sample}
\label{subsec:bzcat}
To search for counterparts to the neutrino data we use the 5BZCat catalog \citep[][]{massaro20155th}.
The catalog is a compilation of 3561 objects where each one has been thoroughly inspected to fulfill the criteria defining a blazar-like nature in 5BZCat (see \origin). The catalog does not rely on any preferential observational energy band or blazar intrinsic physical properties. 

Similarly to \origin, we discard objects that lack an optical spectrum to confirm their blazar nature and appear as ``BL Lac candidate" in the 5BZCat. Then we select objects at $\delta \geq -3^{\circ}$ and located at high latitudes, at $|b| > 10^{\circ}$. The latter selection is motivated by the paucity of objects in the vicinity of the Galactic plane (due to observational biases). As well as because one may not exclude a-priori a contribution by galactic sources to the putative astrophysical signal \citep[][see also Appendix \ref{sec:galplane_hotspots}]{Antares_galactic_ridge:2022}. We remove objects at $  \delta\; > 81^{\circ}$, since no neutrino sky map data are available for regions close to the northern celestial pole. The final sample of 5BZCat blazars employed in the analysis of the northern hemisphere counts a total of 2130 objects. 
For the reader's reference, we summarise in Table \ref{tab:significance} the relevant information for the northern analysis, including those of \origin\ relative to the southern hemisphere analysis.

\section{Statistical analysis procedure}
\label{sec:cross-corr}
In the following we summarise the most salient information about the analysis performed in \originII. For further details we refer the reader to \origin. 
Due to the limited knowledge to guide the selection of the neutrino data, we compute the degree of the blazar/neutrino correlation as a function of two parameters, i.e. the local probability of being astrophysical $L_{\rm min}$ and the association radius $r_{\rm assoc}$. Being these \emph{a posteriori} cuts, a statistical penalty will be accounted for, when estimating the post-trial statistical significance, as explained in the following sections. 

\subsection{Neutrino spot sample} \label{subsec:neutrino_data}
Focusing on the sky map data for $81^{\circ} \geq \delta \geq -3^{\circ}$, we follow the approach of \origin\ and consider only sky positions with the highest $L$ values as it is reasonable to assume that for the majority of the sky locations tested with the likelihood fit, the clustering in the neutrino data may be due to non-astrophysical components such as background fluctuations.
We adopt as the fiducial neutrino-source location (spot) the map-pixel with the highest $L$-value above a pre-defined $L_{min}$. The Right Ascension ($\alpha$) and declination ($\delta$) of the map-pixel are chosen as reference coordinates for the neutrino spot. The separation 
between spots is required to be larger than $1.5^{\circ}$, i.e. larger than the median angular resolution reported for these data \citep{IceCube_10y_reprocessed:2022}. 

Similarly to \origin, we build three subsamples of putative-neutrino sources with decreasing minimum $L$-value ($L_{min}$). In the analysis of the southern hemisphere (see \origin)  the $L_{min}$ ranged between [3.5,4.0,4.5]. In the latest $9$-y north sky map, the $L_{min} \geq 4.5$ subsample contains a small number of spots (only 5 spots are found). Therefore, in the northern analysis we consider the three spots subsamples defined by $L_{min}$ in the range [3.0,3.5,4.0]. These contain 82, 34 and 17 spots, respectively. Consistently with the selection criteria applied to the blazar sample, we select only spots at $|b| > 10^{\circ}$. Upon applying the Galactic plane cut, the final samples of neutrino spots $L3.0$, $L3.5$ and $L4.0$ comprise 66, 29 and 13 spots, respectively.

\subsection{Positional cross-correlation analysis} \label{subsec:cross-correlation}
To estimate the level of correlation between the blazar sample and neutrino data we employ the positional cross-correlation technique presented in \origin\ \citep[see also ][]{Finley:2004,AugerAGN:2008,IceCube_Auger_TA:2016,padovani2016extreme,Resconi2017,plavin2021directional,Hovatta:2021, giommi2020dissecting}.
As association radius for the neutrino spots and blazars we test separation values ($r_{\rm assoc}$) close to the median angular resolution reported for this neutrino dataset, in the range from $1.0^{\circ}$ to $1.4^{\circ}$,  with a step of $0.05^{\circ}$.

We perform a scan in the $\{L_{\rm min},r_{\rm assoc}\}$ space where for each set of $(L^{i}_{\rm min},r^{j}_{\rm assoc})$ we match the positions of the 5BZCat objects with the positions of the neutrino spots. To be conservative, only unique associations are allowed, i.e. one neutrino spot may be matched with only one 5BZCat object at most, and vice versa. This is applied both to the experimental data and to the mock data (see later).
The number of real matches constitutes our test statistics, we denote it as TS$_{\rm astro}^{L_i,r_j}$. The pre-trial p-value, $p^{L_i,r_j}_{\rm pre}$, for the set of parameters is estimated by deriving the chance probability of obtaining a test statistic value equal or higher than the one observed for the real data, following \citet{Davison:1997}. As explained in \origin\ we keep fixed the neutrino spot locations. We simulate $10^6$ Monte Carlo (MC) catalogs by randomizing the blazar positions, preserving both the total number and the spatial distribution of the blazars, applying Kolmogorov-Smirnov tests on the spatial distributions (see \origin). This approach preserves the patterns present in the overall spatial distribution of the blazar sample, at the same time yielding a representative set of random cases. Appendix \ref{appendix:scan} addresses the robustness and impact of the randomization strategy on the results.

%
%
\begin{figure}
    \includegraphics[scale=0.47]{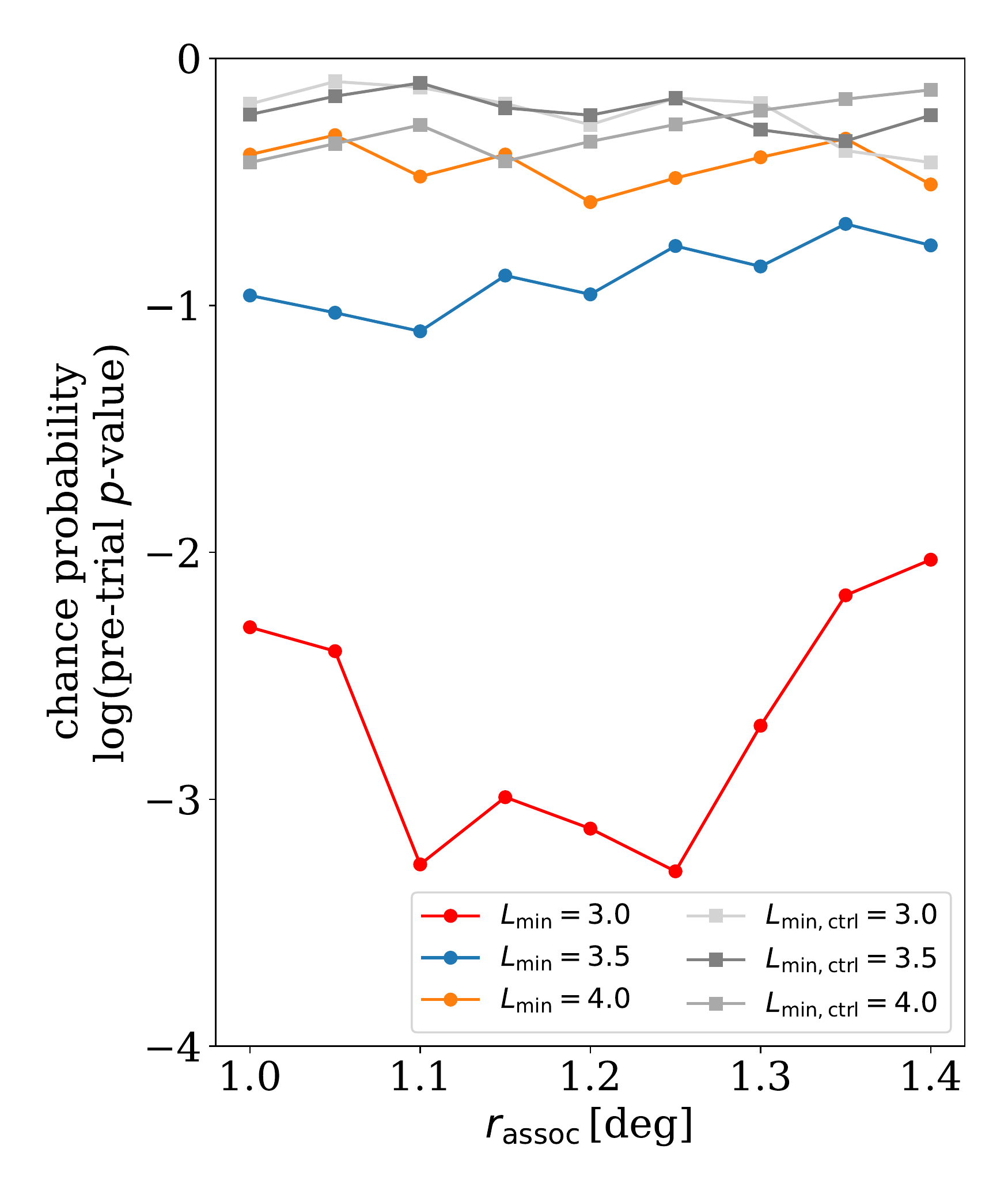} 
        \caption{
        Pre-trial p-value for the 5BZCat blazar/neutrino correlation in the northern sky, as a function of the association radii $r_{\rm assoc}$, for the neutrino data set with $L_{\rm min}=[3.0,3.5,4.0]$ (red, blue, orange circles, respectively). The y-axis displays values in logarithm scale. See Section \ref{sec:cross-corr} for more details. As reference, the results for the northern hemisphere control sample are displayed as gray squares (see Appendix \ref{appendix:control}).
        }
\label{fig:significance}
\end{figure}

In this way, we obtain for each set of $(L^{i}_{\rm min},r^{j}_{\rm assoc})$ a corresponding pre-trial p-value, $p^{L_i,r_j}$ as shown in Figure \ref{fig:significance}. The lowest pre-trial p-value indicates the highest level of correlation potentially present in the data. 
In the northern sky map analysis this corresponds to a  pre-trial p-value, $p^{\rm best}_{\rm pre}=5.12\times 10^{-4}$, observed for $L_{\rm min}=3.0$ and $r_{\rm assoc}=1.25^{\circ}$. 
Since these optimal values are derived from the data, a statistical penalty has to be included in the calculation of the final chance probability, i.e. the post-trial p-value. To this aim, we follow the procedure of \origin\ and obtain a post-trial p-value $p^{\rm best}_{\rm post}=6.79\times 10^{-3}$. The neutrino sub-sample defined by $L \geq 3.0$ contains a set of 66 neutrino spots and, given the employed samples, represents the best trade off between total number of neutrino spots in the sample, and neutrino spots possibly originated by 5BZCat blazars. Throughout \originII, we refer to the northern neutrino spots with $L \geq 3.0$ as hotspots. 42 hotspots have a 5BZCat association within $1.25^{\circ}$, they are listed in Table \ref{tab:associations}.

The robustness of the analysis is tested on a control sample, i.e. astrophysical sources for which no correlation with astrophysical neutrinos is expected. For the northern hemisphere control-sample analysis a post-trial p-value of 1.0 is found. 
For reference, the test was carried out also for the 7-y southern hemisphere analysis, yielding the same results (post-trial p-value of 1.0), although not discussed in \origin\ as the manuscript was yet rich in information. We provide these additional tests in Appendix \ref{appendix:control}, and overlay the control sample results in Figure \ref{fig:significance} (gray points).
%

%
\begin{table*}[ht!] 
\begin{center}
\caption{Statistical significance of the cross-correlation analysis between 5BZCat blazar/neutrino hotspots, performed at high galactic latitudes ($|b| > 10^{\circ}$) . The north analysis is presented in this paper (\originII, Section \ref{sec:cross-corr}), the south analysis in \origin.
The lower part of the table presents the results of the chance probability for the combined north/south-sky experiment ($p^{\rm global}$), estimated using the Fisher's method (see Section \ref{sec:fisher}). The post-trial p-value reported for the northern and southern sky incorporates the effect of testing several data sets. The corresponding significance in Gaussian standard equivalent (one-side) is reported in brackets.
}\label{tab:significance}
\begin{tabular}{@{}cccccccc@{}}
\toprule
Sky region & Dataset (energies) & 5BZCat & Hotspots & Matches  & Pre-trial p-value  & Post-trial p-value & Reference\\
\hhline{========}
North  & $9$~yr data  & 2130 & 66  &  42  & $5.12 \times 10^{-4}$ ($3.28\sigma$) &   $6.79\times 10^{-3}$ ($2.47\sigma$) & \originII \\ 
($-3^{\circ} \leq \delta \leq 81^{\circ}$) & ($\sim$TeV/$\lesssim0.1$~PeV) &&&&&& {\footnotesize Section \ref{subsec:cross-correlation} } \\
\hline  
South  & $7$~yr data & 1177 & 19  & 10  & $3 \times 10^{-7}$ ($4.99\sigma$) &   $2\times 10^{-6}$  ($4.5\sigma$) & \origin  \\  
($-85^{\circ} < \delta < -5^{\circ}$) & ($\gtrsim0.1$~PeV) &&&&&& \\
\midrule \midrule
& & &&  & $p^{\rm global}_{\rm pre}$  & $p^{\rm global}_{\rm post}$ &\\
\hline
North + South &  & -- & -- & -- &  $3.62 \times 10^{-9}$ ($5.78\sigma$) & $2.59\times 10^{-7}$ ($5.02\sigma$)& {\footnotesize Section \ref{sec:fisher}} \\  
\bottomrule
\end{tabular}
\end{center}
\end{table*}
\subsection{Combining northern / southern sky results}
\label{sec:fisher}
The p-values obtained for the northern hemisphere and the southern hemisphere analysis presented in \origin\ may be combined to estimate the chance coincidence probability for the IceCube neutrino data and the 5BZCat catalog global experiment. The two data samples are not correlated and rather complementary, and aim testing the same hypothesis. Hence, the p-values can be combined with the Fisher's method \citep{Fisher:1932}.

Following the Fisher's method, we combine the two pre-trial p-values, i.e. $p^{\rm best}_{\rm pre}=5.12\times 10^{-3}$ for the northern data set, and $p^{\rm best}_{\rm pre}=3\times 10^{-7}$  for the southern hemisphere analysis presented in \origin, and find a global pre-trial p-value $p^{\rm global}_{\rm pre}=3.62\times 10^{-9}$. The global pre-trial p-value represents the strongest statistical correlation potentially present between the blazar catalog and the neutrino data. 

Then, we may combine the two post-trial p-values, and obtain the global post-trial p-value on the experiment. Combining the post-trial p-values estimated for the northern hemisphere, i.e. $p^{\rm best}_{\rm post}=6.79\times 10^{-3}$, and the one for the southern hemisphere analysis presented in \origin, i.e. $p^{\rm best}_{\rm post}=2\times 10^{-6}$, yields a global post-trial p-value of $p^{\rm global}_{\rm post} =2.59 \times 10^{-7}$.
Table \ref{tab:significance} summarises these findings.

Based on the northern hemisphere analysis, the blazar/neutrino correlation holds for the complementary  muon-proxy energy range $\sim$~TeV to $\lesssim0.1$~PeV, corroborating to a higher statistical significance the previous findings. I.e., the observed blazar/neutrino correlation is highly unlikely to arise by chance.  

\section{Previous all-sky searches and neutrino sky maps limitations}\label{sec:previous_searches}
Previous studies that employed the neutrino sky map data followed mostly a ``blind” approach, with the goal of testing whether any of the $\gtrsim 10^7$ individual sky locations shows an inconsistency with background expectations. These searches have the advantage of not requiring any a-priori assumption on the astrophysical signal, at the price of a limited effectiveness due to the large number of trials involved. No evidence for astrophysical point-sources at a statistically significant level ($>5\sigma$) has been reported so far from those searches \citep[see e.g. the latest work in this direction by][]{IceCube_10y_reprocessed:2022}. Nevertheless, they do not rule out the potential presence of genuine astrophysical signals, 
when different hypothesis are tested (see also Appendix B in \origin). Contextually, our study reveals long-term ($\gtrsim$~year-long) astrophysical neutrino emitters, individually below the threshold for firm detection in other previous all-sky searches, that are related to populations of astrophysical sources, i.e. blazars and Seyfert-like objects (see Section \ref{sec:beyond}).

The statistical significance reported in our work may be regarded as a conservative estimate, noting that the northern sky map has a lower spatial resolution ($\sim 0.2^\circ$) compared to the southern one ($\sim 0.1^\circ$). This implies not only a limited accuracy on the position of the putative neutrino sources (spots), it also leads to test a larger number of association radii (e.g. compared to \origin). The latter impacts the experiment by reducing the post-trial statistical significance, as the tested radii count as trials.
However, the association radius is ultimately a proxy for the instrument point spread function (PSF). 
In the presence of a true spatial correlation, one expects (a-priori) a larger number of matches when sampling angular scales that are comparable to the instrument's PSF.

\section{PEVATRON BLAZARS HOSTED IN THE
NORTHERN NEUTRINO SKY}\label{sec:pevatron_blazars}

%
%
\begin{figure*}[t!]
    \centering
    \includegraphics[scale=0.38]{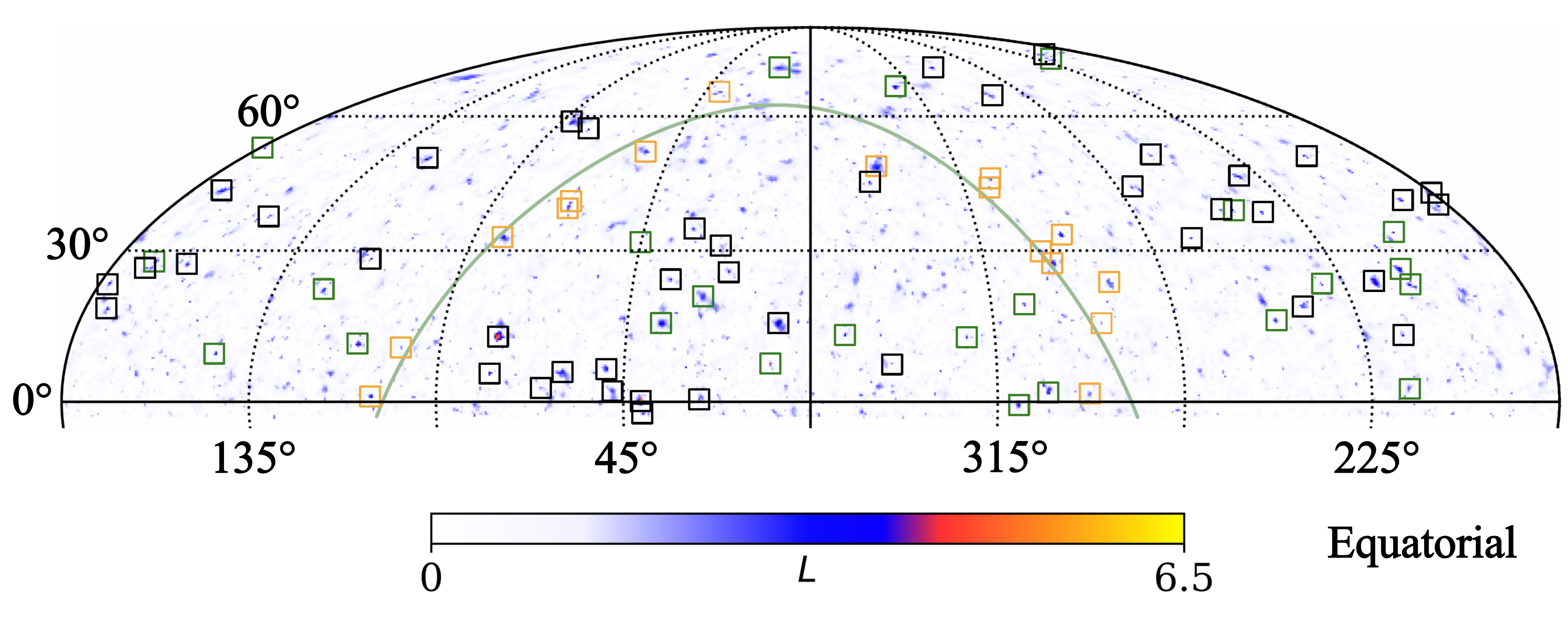}
    \caption{
    Sky map in equatorial coordinates (J2000) of the IceCube neutrino local p-value logarithms denoted as $L$. Locations of PeVatron blazars associated with neutrino hotspots are pointed out by black squares. Unassociated hotspots are highlighted by green squares. Hotspots that are located along the galactic plane and, hence were not included in the statistical analysis, are highlighted by orange squares. 
    Squares are not to scale and serve the only purpose of highlighting  hotspots' locations. %
    The Galactic plane is shown for reference as a green line. No smoothing is applied to the IceCube sky map.}
    \label{fig:north-sky}
\end{figure*}%

Figure \ref{fig:north-sky} displays the $\sim9$-y northern-sky IceCube map in equatorial coordinates. The sky locations of the neutrino hotspots with a 5BZCat blazar association, i.e. PeVatron blazar, are highlighted by black squares. Hotspots that remain unassociated are reported as green squares, while hotspots located close to the Galactic plane are shown as orange squares (see Appendix \ref{sec:galplane_hotspots}). The samples are also listed in Table \ref{tab:associations}, Table \ref{tab:unassociated_spots} and Table \ref{tab:galplane_hotspots}, respectively.
We observe a roughly isotropic distribution of neutrino hotspots across the northern sky.

Among the 42 blazar candidate associations with neutrino hotspots found in the northern hemisphere analysis, 19 have a counterpart in the latest $Fermi$-LAT source catalog \citep[see Table \ref{tab:associations}, ][]{4FGL_DR3:2022}, and, therefore, are $\gamma$-ray emitters at MeV-GeV energies. Only 10 were listed in the 2nd $Fermi$-LAT AGN Catalog \citep[2LAC,][]{2LAC} and, hence, included in the neutrino stacking analysis of \citet{IceCube2017_2LAC}. In line with the findings of \origin, PeVatron blazars  cover a fairly large range of redshifts, and are in general weak $\gamma$-ray emitters. A large number of them remain undetected at GeV $\gamma$-ray energies suggesting that, overall, in blazars the neutrino emission is not necessary related to the observed $\gamma$-ray emission.

\section{Northern \& southern \\ PeVatron Hotspots}\label{sec:pev_events}
%
In this section we briefly discuss further evidences supporting the hypothesis that the observed hotspots may be driven by astrophysical sources capable of producing neutrinos with energies up to PeVs.

\subsection{Southern Hotspots \& Glashow resonance event}\label{subsec:glashow}
%
A shower of particles consistent with being produced by a Glashow resonance has been recorded by the IceCube Observatory on 2016 December 8. It originates from the southern celestial hemisphere, and is found to be consistent with a W$^{-}$ boson that decays hadronically, and has an estimated energy of $6.05 \pm 0.72$~PeV \citep{IceCube_glashow:2021}. The neutrino event is very likely to be of astrophysical origin, and has a best-reconstructed position of $\alpha=192.70^{\circ}$ and $\delta = -15.15^{\circ}$. Because it was observed in 2016, it is not included in the data used to produce the 7-y IceCube sky map studied in \origin, that encompasses timescales between 2008-2015. 

\begin{figure*}[t!]
  \centering
    \includegraphics[scale=0.56]{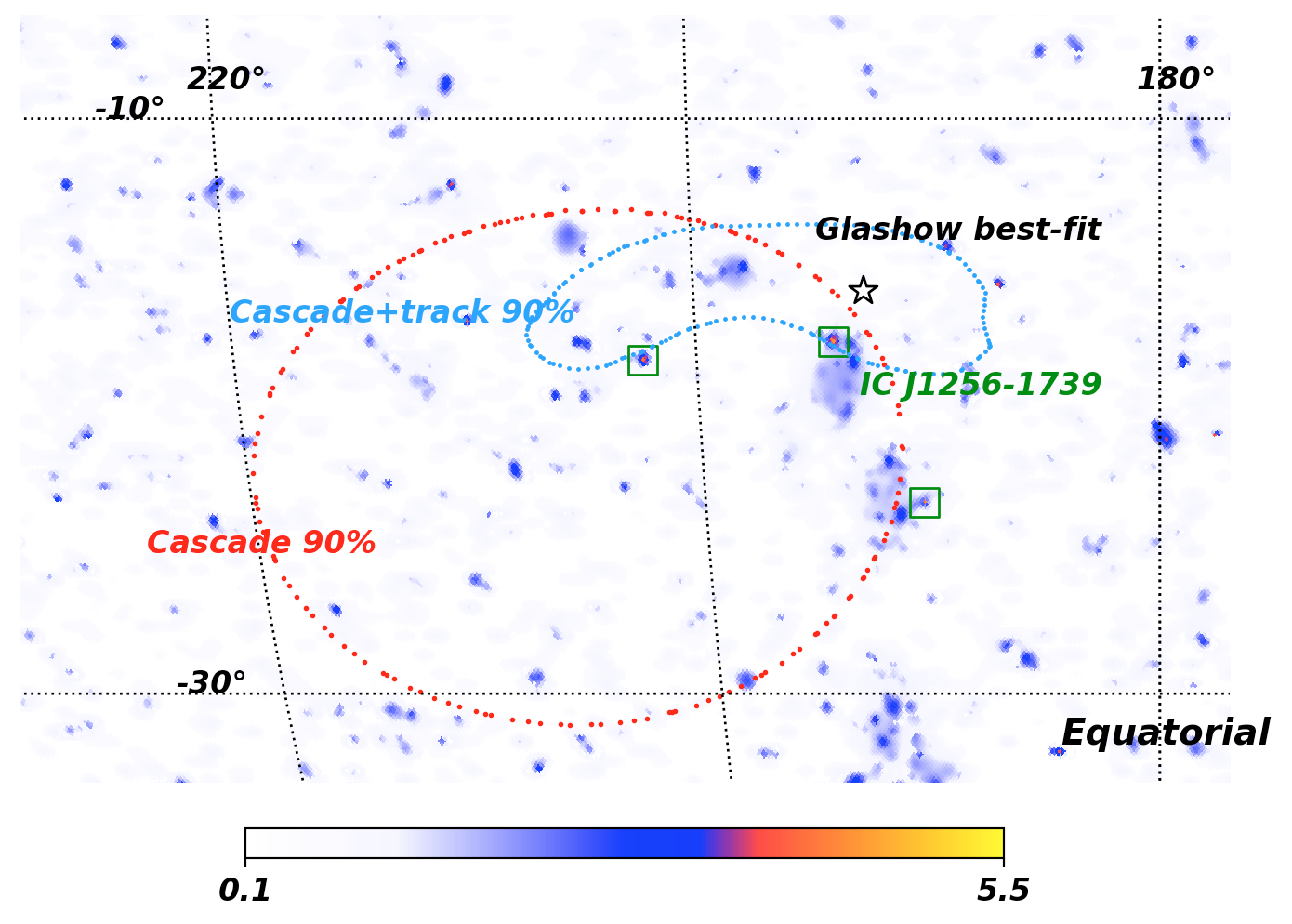}
    \caption{Glashow resonance event \citep{IC_glashow:2021,IceCube_glashow:2021} discussed in Section \ref{subsec:glashow}.
    The most-probable direction of the PeV event is shown as a black star. The blue and red contours indicate the 90\% positional reconstruction derived with two different methodologies. For more details we refer the reader to \citet{IceCube_glashow:2021}. In the foreground is the 7-y sky map (see \origin), with the hotspots pinpointed by our work highlighted by green squares. No smoothing is applied to the IceCube sky map.
    }
    \label{fig:glashow}
  \end{figure*}

Figure \ref{fig:glashow} displays the 7-y sky map with overlaid the positional reconstruction region of the Glashow resonance event. Three hotspots found in \origin\ are located in the vicinity of this PeV event. In particular, the unassociated hotspot IC~J1256$-$1739 is $2.19^{\circ}$ away from the best-fit position of the Glashow resonance event, and located within its $90\%$ positional reconstruction contour. 

An enhanced probability of observing electron antineutrinos from blazars, induced by the Glashow resonance, was anticipated by \citet[][]{Mannheim1993} in the seminal work of the ``proton blazar'' model.
The 5BZCat object 5BZQ~J1258$-$1800 is located $0.66^{\circ}$ away from IC~J1256$-$1739.
Although this separation does not allow us to claim a confident association, we suggest that this blazar may be a candidate counterpart for the hotspot and, possibly, the Glashow event. The separation is close to the optimal association radius found for the southern sky analysis, i.e. $r_{assoc}=0.55^{\circ}$ (see \origin), especially when accounting for the $0.1^{\circ}$ resolution of the 7-y sky map. 

\subsection{Northern PeV neutrino muon-track \\ consistent with IC~J0721$+$1125}
%
On 2014 June 11 a PeV track-like event was recorded by the IceCube Observatory \citep{IceCube_Schoenen_PeV_event:2015}. The probability of this event being of atmospheric origin has been estimated to be less than 0.005\%, strongly suggesting an astrophysical origin \citep{IceCube_north_6y:2016}.
The deposited energy has been measured to be $2.6\pm0.3$~PeV of equivalent electromagnetic energy,
with a reconstructed median muon-proxy energy of $4.45\pm1.2$~PeV (68\% C.L). The median expected muon-neutrino energy is 8.7~PeV.
 The PeV track-like event has reconstructed celestial coordinates $\alpha= 110.65^\circ \, (^{+0.53^\circ}_{-0.62^\circ}), \delta=11.45^\circ ({\pm 0.19^\circ})$ ($90\%$ C.L.) \citep{IC_track_alert_aclog:2023}
 consistent with the unassociated hotspot IC~J0721$+$1125, located at $\alpha= 110.21^\circ, \delta=11.42^\circ$. No 5BZCat object is positionally consistent with this hotspot. The lack of a candidate counterpart to this promising neutrino hotspot may be due to the blazar catalog incompleteness, hence related to the fainter and/or far away population of AGN out of reach of current electromagnetic surveys, or may as well be ascribable to a different population of neutrino emitters.

\section{Beyond the multi-messenger \\ blazar / neutrino connection}\label{sec:beyond}
The discovery of PeVatron blazars builds on the theoretical prediction that if jetted AGN, in particular those of the blazar class, accelerate hadrons, the accompanying neutrino emission may be observable at the energies accessible by the IceCube observatory. The investigative approach presented in \origin\ and \originII, 
offers a new view to decode the current and future neutrino data and further exploit them in the multi-messenger context. Assuming that neutrino hotspots are driven by genuine astrophysical sources, our findings suggest additional populations of cosmic neutrino emitters besides blazars, noting that about half of the neutrino hotspots remain unassociated, and/or multiple mechanisms for neutrino production as briefly discussed in the following.

\subsection{Jetted active galaxies in the local Universe}
Amongst the objects associated with high-energy neutrinos, some stand out in the ``enlarged'' blazar family of 5BZCat given their radio characteristics that depart from the general blazar fingerprint, i.e. a single compact radio component typically interpreted as a relativistic jet seen at a small viewing angle. Among others, there are 5BZU~J1647$+$4950, a.k.a. SBS~1646$+$499, located within the association radius of IC~J1644$+$5052, 
and 5BZU~1819$-$6345, a.k.a. PKS~1814$-$63, located within the association radius of IC~J1818$-$6315 (see \origin). 

Despite displaying the typical double-hump spectral energy distribution of a blazar, the former appears analogous to the radio galaxy Centaurus A \citep{Pajdosz_J1647:2018}. The radio morphology of 5BZU~J1647$+$4950 is complex, with giant megaparsecs radio lobes, probably formed during earlier epochs of jet activity. In the vicinity of the central compact core there is indication of recent reactivation of the nuclear jet, 
while the larger-scale jet appears structured, with a spine-boundary shear layer morphology. In its host galaxy, type II core-collapse supernovae occur \citep[e.g. SN 2009fe; CBET 1820,][]{SN2009fe}. 

The latter, i.e. 5BZU~1819$-$6345, is a peculiar powerful radio emitter \citep[][]{ODea:1990,Labiano:2007}, sometimes classified as a radio galaxy. In addition to a central compact radio component, it displays two lobes and, it may host a young radio-jet that is starting to expand through the near-nuclear interstellar medium \citep[ISM,][]{Santoro:2020}. In such case, we might be witnessing an initially relativistic jet that is drilling its way out, and that is rapidly decelerated by collisions in a dense surrounding broad line region or the ISM \citep[][]{Morganti:2011}.
The radio power of 5BZU~1819$-$6345 (P$_{5GHz} \sim 10^{25}$~W~Hz$^{-1}$) is two orders of magnitude higher than the most powerful radio Seyfert galaxy, i.e. NGC~1068 (P$_{5GHz} \sim 10^{23}$~W~Hz$^{-1}$). 
5BZU~1819$-$6345 shares similarities with NGC~1068, that is a nearby ($\sim$14~Mpc) active galaxy suggested as neutrino emitter \citep{IceCube10y:2020,IceCube_10y_reprocessed:2022}, among others an host galaxy morphology dominated by a strong, warped disk component that is observed edge-on.
It has been suggested as an extreme example of the effects of outflow-cloud interactions in galaxies containing a rich interstellar medium, and perhaps a missing link between radio galaxies and radio-loud Seyfert galaxies \citep{Morganti:2011}. 

Both 5BZU~J1647$+$4950 and 5BZU~1819$-$6345 show properties common to Seyfert-like galaxies. They are located in the nearby universe (within $\sim 200$~Mpc and $\sim$280~Mpc, respectively), and present remarkable features, ideal to foster the production of high-energy neutrinos, e.g. presence of AGN-driven outflow and/or corona activity, prominent jet-cloud interaction, among others.

\subsection{Hotspot IC J1210+3939 \& 5BZB~J1210+3929,\\ in the vicinity of NGC 4151}
\origin\ envisaged the possibility that astrophysical neutrino hotspots could be generated by the cumulative emission of multiple sources. For instance, for a given spot, (neutrino-emitter) objects whose spatial separation is smaller than the angular resolution of current neutrino observatories, would seemingly contribute to enhancing the neutrino clustering local p-value.
Further, we anticipated that some of the hotspots could be possibly driven by other populations of neutrino emitters besides blazars, as supported by the detection of hadronic $\gamma$ rays from ultra fast outflows (UFOs) hosted in AGN \citep{UFO:2021}.

A possible example in this direction is highlighted in Figure \ref{fig:field_ngc4151}, which shows a close up view of the $\sim$~9-y sky map containing the hotspot IC~J1210$+$3939, associated with 5BZB~J1210+3929. Hotspots and associated blazars are pinpointed by black squares and crosses, respectively. In the region, there is also the jetted AGN NGC~4151 (magenta circle), located 5\arcmin\ away from 5BZB~J1210+3929, and 0.2$^\circ$ from IC~J1210$+$3939.  NGC~4151, which is not listed in 5BZCat, is a bright Seyfert galaxy suggested as promising neutrino source \citep[e.g.,][]{Stecker_AGN_cores:1991}. Similarly to the other Seyfert-like objects of our PeVatron sample, it is found in the nearby Universe ($\sim 15$~Mpc, $z$=0.0033). 

NGC~4151 is known to host an UFO and is included in the study that provides evidence for collective, hadronic $\gamma$-ray emission from a small set of UFOs \citep{UFO:2021}. According to \citet{UFO:2021}, AGN winds are capable of accelerating cosmic rays with proton maximum energies up to $ 10^{17}$~eV, suggesting that they could contribute to the diffuse astrophysical neutrino flux in the IceCube energy range \citep[e.g.][]{Wang_Loeb_nu_bkg:2016}. 
While UFOs are predicted to be generally too faint to be detected individually by the $Fermi$-LAT at $\gamma$ rays, NGC~4151 showed hints for the overall highest individual $\gamma$-ray emission. A recent study in the literature claims its detection at $\gamma$ rays at high confidence \citep[$\gtrsim 5\sigma$,][]{Peretti:2023}.
If this were to be confirmed, NGC~4151 would represent an outlier Seyfert galaxy based on the observed $\gamma$-ray properties. Moreover, theoretical models that attempt to provide a concomitant explanation of the putative neutrino and $\gamma$-ray emission from NGC~4151 face major challenges. For instance, models based on the UFO and/or coronal component predict neutrino fluxes undetectable by IceCube, given the $Fermi$-LAT spectrum \citep[][]{UFO:2021,Inoue:2021}. 

Caution should be however paid when considering such claims. We note that the analyses of \citet{Peretti:2023} relies on outdated LAT catalogs to model the $\gamma$-ray emission from the region \citep[][i.e. the 8-yr LAT catalog]{abdollahi2020_4FGL}.
An inspection of the most recently released $Fermi$-LAT point-source catalog indicates a new faint $\gamma$-ray source (detected at $\gtrsim5\sigma$) in the vicinity of IC~J1210$+$3939, i.e. 4FGL~J1210.3$+$3928 \citep[][i.e. the 12-yr LAT catalog]{4FGL_DR3:2022}. Although the $\gamma$-ray emission is spatially close to NGC~4151, it is associated with the blazar 5BZB~J1210+3929 \citep[see also the latest LAT AGN catalog, 4LAC-DR3,][]{4LAC_DR3:2022}.
The association with the blazar offers a natural explanation also in light of the observed neutrino emission: 5BZB~J1210+3929 may represent a typical PeVatron blazar, whose comparatively weak GeV emission is emerging with the integration of more LAT exposure.
While at the current state one may not exclude a contribution from NGC~4151 to the observed neutrino and $\gamma$-ray emission, this appears to be minor, unless invoking new physical interpretations.

\section{Discussion and Conclusions}  \label{sec:conclusions}
Conducting the positional cross-correlation analysis between the IceCube northern hemisphere sky map and 5BZCat sources yields a post-trial $p$-value of $6.79\times10^{-3}$. This result is combined with the post-trial $p$-value of $2\times10^{-6}$ for the southern hemisphere, presented in \origin, following the Fisher's method. The global post-trial $p$-value for the chance association between IceCube neutrino hotspots and 5BZCat blazars is found to be at the level of $2.59\times10^{-7}$. 

Based on this work, blazars are the first population of extragalactic neutrino sources discovered at high confidence. The two complementary studies point out 52 PeVatron blazars as likely counterparts of high-energy neutrinos, along with 23 hotspots missing an astrophysical counterpart in the 5BZCat. Only 22 PeVatron blazars have been detected at GeV energies by the $Fermi$-LAT \citep[see Table \ref{tab:associations}, ][]{4FGL_DR3:2022}, confirming the conclusions of \origin. 

The majority of PeVatron blazars displays a radio power P$_{1.4GHz} \gtrsim 10^{26}$~W~Hz$^{-1}$, which is typical of high-excitation radio galaxies \citep[HERGs,][]{Best_Heckman:2012}, pushing forward as preferred neutrino emitters objects equipped with radiation fields external to the jet (e.g., from the accretion disc, coronae, photons reprocessed in the broad-line region and from the dusty torus).
On the other hand, objects with properties that depart from the ordinary definition of blazars among the PeVatron sample (see Section \ref{sec:beyond}), along with NGC~1068, could represent the first prototypes of a more abundant and (relatively) faint population of neutrino emitters in the nearby universe. Notably, HERGs objects have emission-line spectra similar to some Seyfert galaxies.
Seyfert-like galaxies with outflows and/or jets are suggested to be active galaxies in their early stage of radio activity and, as such, are likely to have significant coronal activity, with their nuclear regions wrapped in a cocoon or thick disk of material remnant from the event(s) that trigger the activity. The jet expanding in this medium interacts and sweeps out the rich ISM, providing a good environment for neutrino production.

For objects with a significant coronal activity (as proposed for NGC~1068), in the presence of highly magnetized and turbulent coronae, cosmic rays can be accelerated and subsequently depleted via hadronuclear and photohadronic interactions, providing targets for $\sim1-100$~TeV energy neutrino production. Among the various acceleration processes, capable to accelerate cosmic rays in the corona environment, the magnetic reconnection is the one reaching the highest neutrino energies with values above 100~TeV \citep[][ and references therein]{Kheirandish:2021}. Even though the magnetic reconnection has lower flux normalization compared to the stochastic cosmic ray pressure process, still this scenario could provide promising primers to guide future studies.

\begin{acknowledgments}
This work was supported by the European Research Council, ERC Starting grant \emph{MessMapp}, S.B. Principal Investigator, under contract no. 949555. 
The authors are grateful for valuable conversation to M. Santander, J. DeLaunay. 
This work has made use of data from the Space Science Data Center (SSDC), a facility of the Italian Space Agency (ASI), and data provided by the IceCube Observatory.
\end{acknowledgments}

%

\vspace{5mm}
\facilities{The IceCube Observatory, The Fermi-Large Area Telescope}


\software{Astropy \citep{2013A&A...558A..33A,2018AJ....156..123A},  
          Healpy \citep{Zonca:2019}, HEALPix \citep{Gorski:2005}, Topcat \citep{Taylor:2005}.
          }



\appendix

\begin{figure*}[t!]
  \centering
    \includegraphics[scale = .8]{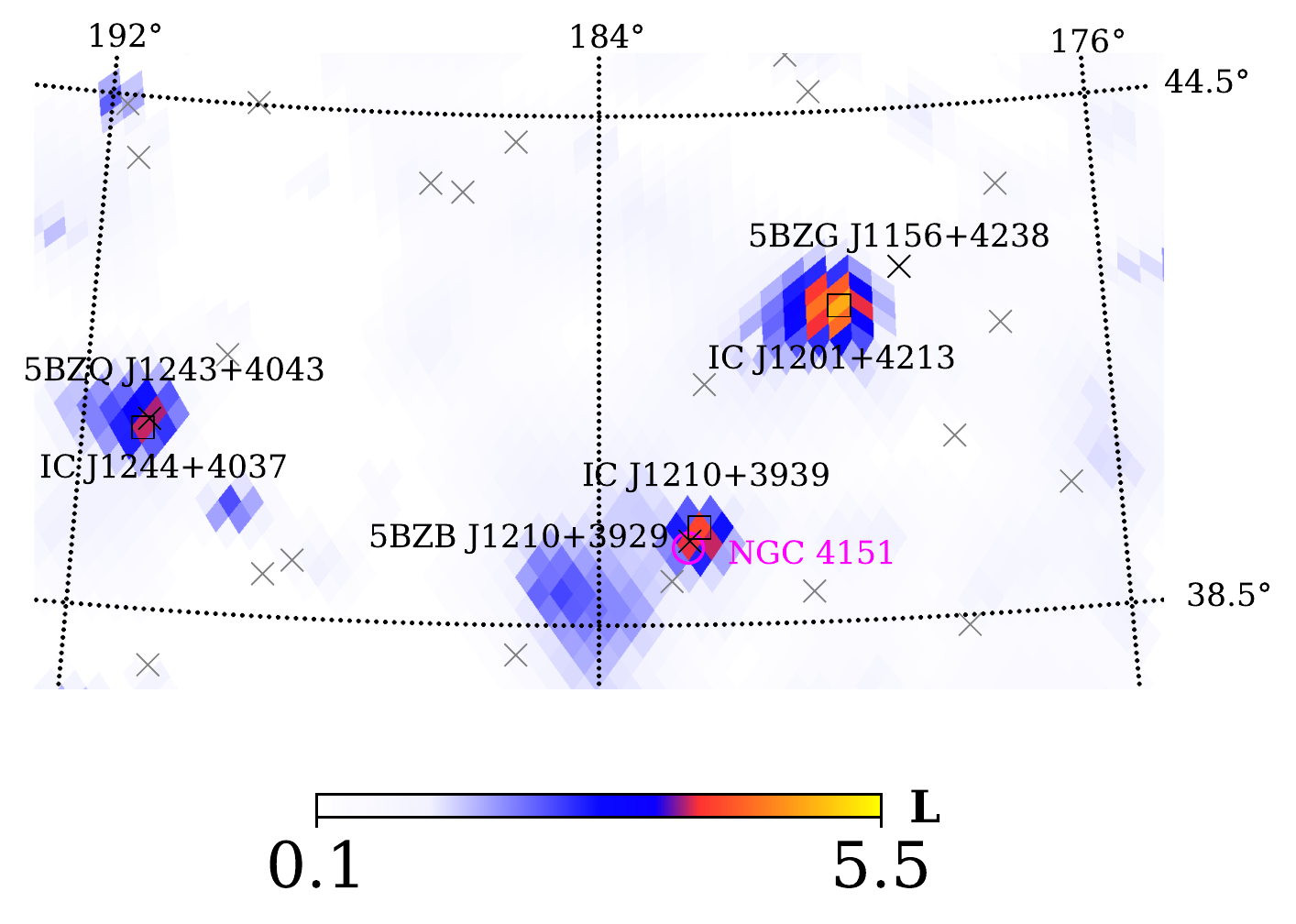}
    \caption{
Close up view of the IceCube sky map, in equatorial coordinates, for the region containing the hotspot IC~J1210$+$3939, associated with 5BZB~J1210+3929. Locations of hotspots and associated blazars are highlighted by black squares and crosses, respectively. Gray crosses display 5BZCat in the field. The magenta circle indicates NGC~4151, located 5\arcmin\ away from 5BZB~J1210$+$3929, and 0.2$^\circ$ from IC~J1210$+$3939. The coordinate grid has $8^\circ \times 6^\circ$ steps. No smoothing is applied to the IceCube sky map.
    }
    \label{fig:field_ngc4151}
  \end{figure*}

\section{Statistical analysis robustness}\label{appendix:scan}
\begin{figure}[h!]
  \centering
  \begin{minipage}{.45\linewidth}
    \includegraphics[scale=0.4]{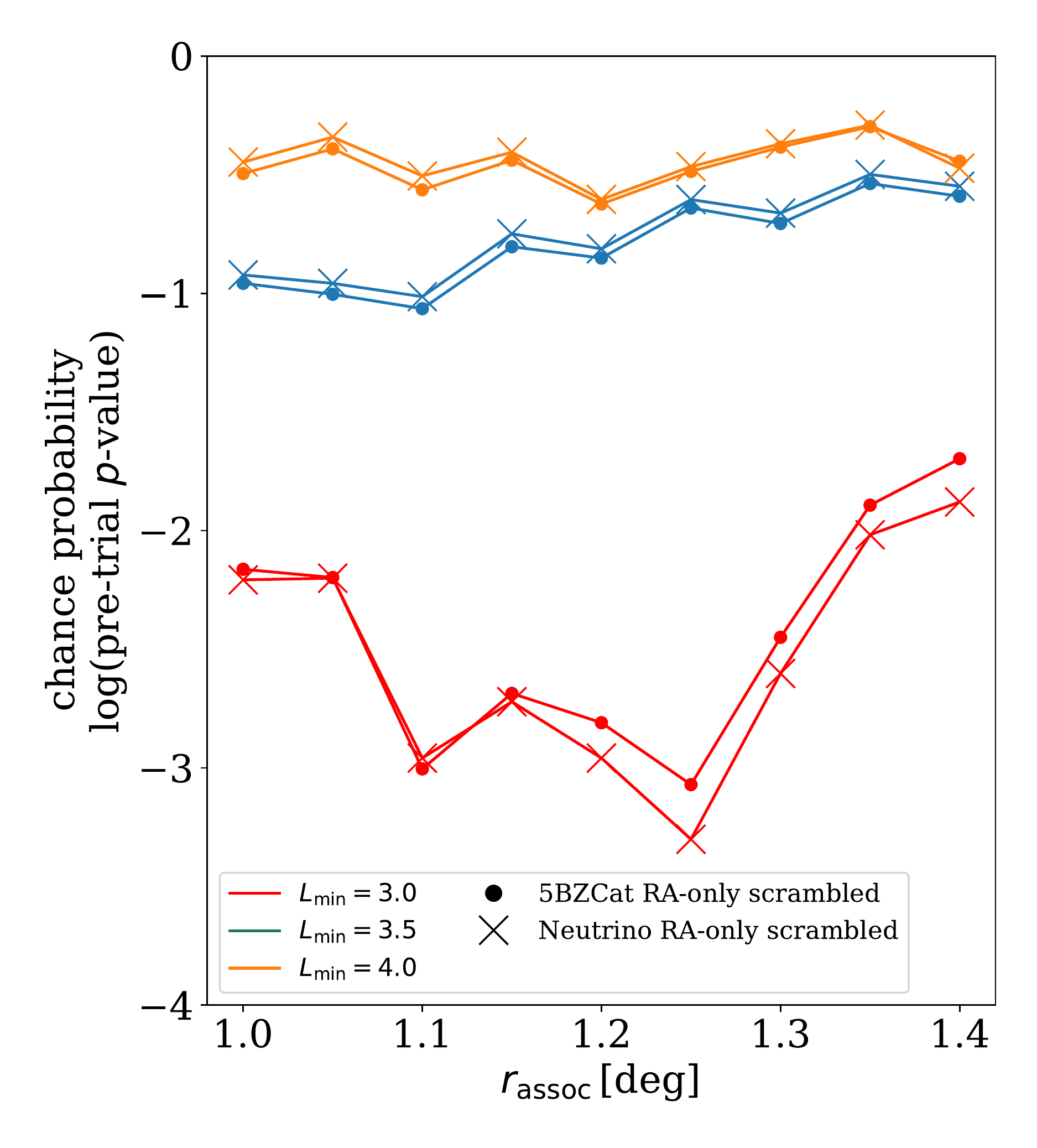} 
     \end{minipage}
       \begin{minipage}{.4\linewidth}
    \captionof{figure}{
    Same as Figure \ref{fig:significance} but obtained by randomising the neutrino spot positions only in Right Ascension (crosses), and the 5BZCat objects only in Right Ascension (filled circles). See Appendix \ref{appendix:scan}. 
    The y-axis displays values in logarithm scale.  
        \label{fig:HS_ra}
    }\end{minipage}
  \end{figure}

As a-posteriori robustness test we repeated the cross-correlation analysis by keeping the blazar sky distribution fixed and randomizing the IceCube neutrino spots. This was done by randomizing the right ascension of the neutrino spots while keeping their declination fixed, thus accounting for a potential residual declination dependence (due to the IceCube acceptance), and repeating the steps of the statistical analysis described in the main text. 
This procedure yields a minimum pre-trial p-value of $6.00 \times 10^{-4}$ which is observed for $L_{\rm min}=3.0$ and $r_{\rm assoc}=1.25^{\circ}$, and a post-trial p-value of $7.10 \times 10^{-3}$ (see Figure \ref{fig:HS_ra}, crosses). A slighter higher post-trial p-value with respect to the one in Table \ref{tab:significance} is expected because the randomization is applied only to the right ascension. As a matter of fact, when applying (a-posteriori) the randomization to the blazar right ascension only,
we obtain comparable results (filled circles in Figure \ref{fig:HS_ra}), i.e. minimum pre-trial p-value of $8.60 \times 10^{-4}$ and a post-trial p-value of $1.15 \times 10^{-2}$.
%
%
%
%
  
\section{Control sample}\label{appendix:control}
The statistical analysis employed in the cross-correlation between 5BZCat and the northern (Section \ref{sec:cross-corr}), and southern hemisphere neutrino data (\origin), was applied to a control sample, i.e. objects which are not expected to be neutrino emitters and thus should display no spatial correlation with the neutrino hotspots. To this aim, we employ sources which are not blazars. An established fingerprint for blazar activity is a radio flux density $\gtrsim$ tens of mJy as observed in the $\sim 1$~GHz band. Therefore, we select faint radio sources, i.e. with a flux density $<3$~mJy, chosen randomly from the NRAO VLA Sky Survey (NVSS) at 1.4~GHz \citep{NVSS_survey:1998}  and the Sydney University Molonglo Sky Survey (SUMSS) survey at 843~MHz \citep{SUMSS_survey:2003}. These two surveys include more than a hundred thousand objects and, combined, cover the full sky. 
The control sample is built from NVSS between $81^{\circ} \geq \delta \geq -40^{\circ}$ and SUMSS between $-85^{\circ} < \delta < -40^{\circ}$. We avoid objects at galactic latitudes $|b| \leq 10^{\circ}$, and ensure that the total number of sources considered in the cross-correlation analysis is the same as in the respective 5BZCat samples, to be consistent with the experiment on the real data. The northern control sample is composed by 2130 objects located between $81^{\circ} \geq \delta \geq -3^{\circ}$, and at $|b| > 10^{\circ}$. The southern control sample counts 1177 objects between $-85^{\circ} < \delta < -5^{\circ}$, and at $|b| > 10^{\circ}$.

We perform the statistical cross-correlation analysis described in Section \ref{sec:cross-corr} between the northern control sample and the neutrino spots samples employing $10^{5}$~MCs.
For the northern hemisphere sample the best pre-trial p-value is $0.4$, found for the set of parameters $L_{\rm min}=4.0$ and $r_{\rm assoc}=1.00^{\circ}$; accounting for trials leads to a post-trial p-value of 1.0. Figure \ref{fig:significance} displays the results of the $\{L_{\rm min},r_{\rm assoc}\}$ space scan for the northern control sample (gray), along with the results for the 5BZCat northern analysis (orange, blue, and red).
We provide for reference also the results obtained by applying the analysis described in \origin\ to the southern hemisphere control sample. The best pre-trial p-value is $0.1$, found for the set $L_{\rm min}=3.5$ and $r_{\rm assoc}=0.45^{\circ}$,
correcting for trials yields a post-trial p-value of 1.0. 

\section{Low-galactic latitude neutrino hotspots}\label{sec:galplane_hotspots}
%
Neutrino hotspots located close to the the Galactic plane ($|b| \leq 10^{\circ}$) were not included in the cross-correlation analysis performed in the northern (Sect. \ref{subsec:cross-correlation}) and southern hemisphere (\origin). The exclusion of these hotspots is mainly motivated by the poor completeness of the blazar catalog closer to the Galactic plane region. Besides, along the Galactic plane, one may not exclude a priori a possible galactic origin for these neutrino hotspots. 
In terms of putative neutrino astrophysical genuineness, however, there is no reason to prefer higher galactic latitude hotspots to the lower galactic latitude ones. In principle, these low-galactic latitude hotspots may be promising candidate neutrino sources. 
As reprove of this, 
evidence ($\gtrsim4\sigma$) for neutrino emission from the Galactic plane has been claimed\footnote{The IceCube collaboration presented this finding at the 2023 Deutsche Physikalische Gesellschaft Spring meeting, and the 2023 American Physical Society Spring meeting.} by the IceCube collaboration, based on an improved IceCube dataset \citep[see also][]{Antares_galactic_ridge:2022}.

In the northern hemisphere sample, 16 hotspots are close to the Galactic plane, at $|b| \leq 10^{\circ}$. A-posteriori we note that two of them have a 5BZCat object located within the association radius found in this work (5BZU~J1911$+$1611 and 5BZQ~J0228$+$6721). For reference, they are listed in Table \ref{tab:galplane_hotspots} along with the celestial coordinates and their corresponding $L$ values from the IceCube sky maps. In the Table we report also the galactic hotspots found in the southern hemisphere sample analysed in \origin.

\section{A-posteriori comparison to previous neutrino/blazars correlation studies}
An analysis presented in \citet{plavin2021directional} investigates the possible correlation between IceCube neutrinos and  a sample of 3411 radio-bright AGN, defined by an 8~GHz flux density $S_{8GHz} \geq 150$ mJy. The authors apply a time-integrated likelihood technique to compute the significance of the directional correlation between radio-selected objects and a list of candidate neutrino locations. The latter list is a combination of IceCube  high-energy neutrino-candidate events and sky locations based on the same 7-years IceCube data studied in \origin.
The radio-selected sample contains 1938  sources at $\delta > -5^{\circ}$, and is correlated with all the putative neutrino locations of the northern sky map ($\delta > -5^{\circ}$). The study reports a chance association between the radio sources and the IceCube data at the $3.0\times10^{-3}$ level, when taking into account the radio-brightness of the objects. The radio-bright selected sample 
has a total of 1579 objects located at $81^{\circ} \geq \delta \geq -3^{\circ}$ and $|b|>10^{\circ}$; 952 of them are objects in common with the 5BZCat northern sample presented in Section \ref{subsec:bzcat}. We perform a-posteriori the statistical analysis described in Section \ref{sec:cross-corr} with the radio-selected sample and the northern neutrino spot samples of Section \ref{sec:samples}, employing $10^{5}$~MCs.  
The best pre-trial p-value is $2 \times 10^{-2}$, observed for the set of $L_{\rm min}=3.5$ and $r_{\rm assoc}=1.10^{\circ}$. Accounting for trials yields a post-trial p-value of $3 \times 10^{-1}$, for a chance correlation between the radio-bright selected sample and the neutrino hotspots.

%
%
\begin{table*}
\begin{center}
\caption{List of high-galactic latitude ($|b| > 10^\circ$) neutrino hotspots / PeVatron blazars associations found by this study. The columns report the neutrino hotspots name, equatorial coordinates (J2000) and the $L$ value of the hotspot. The candidate 5BZCat blazar counterpart along with the redshift and the distance between the blazar and hotspot.}\label{tab:associations}
\begin{threeparttable}
\resizebox{0.8\textwidth}{!}{ 
\begin{tabular}{@{}lrcclcc@{}}
\toprule
IceCube hotspots &&&& Blazar associations&\\
\hhline{=======}
&   $\alpha_{hs}$[$^{\circ}$]   &   $\delta_{hs}$[$^{\circ}$]   & $L$ & 5BZCat & $z$ & Separation[$^{\circ}$] \\ 
\cline{1-7}
IC J0242$-$0214	&	40.43	&	$-$2.24	& 3.827	    & 5BZQ J0239$-$0234	& 1.11413$^{A}$	&	0.60	\\
IC J0243$+$0009	&	40.78	& 0.15	    & 6.752	    & 5BZB J0243$+$0046$^{a}$	& 0.409         &	0.63	\\
IC J0147$+$0027	&	26.72	&	0.45	& 3.020	    & 5BZB J0148$+$0129$^{a}$	& 0.94206$^{B}$	&	1.12	\\
IC J0311$+$0205	&	47.64	&	2.09	& 3.028	    & 5BZQ J0312$+$0133$^{a}$	& 0.664      	&	0.76	\\
IC J0419$+$0241	&	64.86	&	2.69	& 3.155	    & 5BZQ J0422$+$0219$^{a}$	& 2.277      	&	0.93	\\
IC J0509$+$0532	&	77.34	&	5.53	& 4.125	    & 5BZB J0509$+$0541$^{a,b}$	& 0.3365$^{C}$	&	0.16	\\
IC J0359$+$0550	&	59.77	&	5.83	& 3.413	    & 5BZQ J0400$+$0550	&	0.758$^{D}$	&	0.28	\\
IC J0317$+$0626	&	49.22	&	6.43	& 3.334	    & 5BZB J0314$+$0619$^{a}$	&	0.62$^{E}$	&	0.62	\\
IC J2241$+$0720	&	340.31	&	7.33	& 3.265	    & 5BZQ J2238$+$0724	&	$>$1.011	&	0.77	\\
IC J0505$+$1247	&	76.29	&	12.79	& 5.192	    & 5BZB J0502$+$1338$^{a}$ & 0.35$^{F}$	&	1.06	\\
IC J1421$+$1306	&	215.16	& 13.09	    & 3.444	    & 5BZQ J1420$+$1205	& 4.026         &	1.00	\\
IC J0032$+$1524	&	7.91	&	15.40	&	3.667	& 5BZB J0035$+$1515$^{a}$	& --	        &	0.88	\\
IC J1139$+$1822	&	174.73	& 18.37	    &	3.379	& 5BZQ J1143$+$1843	& 0.37435$^{B}$ &	1.06	\\
IC J1551$+$1841	&	237.66	& 18.68	    &	3.291	& 5BZB J1546$+$1817$^{a}$	& --            &	1.08	\\
IC J1152$+$2309	&	177.89	&	23.16	&	4.317	& 5BZB J1150$+$2417$^{a}$	& 0.209$^{F}$	&	1.18	\\
IC J1428$+$2348	&	216.91	& 23.81	    & 4.179	    & 5BZB J1427$+$2348$^{a,c}$	& 0.604$^{G}$   &	0.15	\\
IC J0222$+$2408	&	35.51	&	24.13	& 3.107	    & 5BZU J0220$+$2509	& 0.48$^{H}$	&	1.06	\\
IC J0124$+$2537	&	20.92	&	25.61	& 3.772	    & 5BZQ J0122$+$2502	& 2.025     	&	0.62	\\
IC J1125$+$2627	&	171.21	&	26.44	& 3.440	    & 5BZQ J1125$+$2610	& 2.35005$^{B}$	&	0.36	\\
IC J1045$+$2717	&	161.19	&	27.28	& 3.175	    & 5BZQ J1047$+$2635	& 2.56364$^{B}$ &	0.94	\\
IC J0738$+$2817	&	114.43	&	28.29	& 3.352	    & 5BZB J0737$+$2846	& 0.272        	&	0.51	\\
IC J0135$+$3102	&	23.73	&	31.04	& 3.167	    & 5BZQ J0137$+$3122	& 1.73063$^{B}$	&	0.58	\\
IC J1712$+$3237	&	258.05	&	32.62	& 3.088	    & 5BZU J1706$+$3214	& 1.06929$^{B}$	&	1.20	\\
IC J0206$+$3435	&	31.46	&	34.59	&	3.133	& 5BZB J0208$+$3523$^{a}$	&	0.318     	&	0.98	\\
IC J1000$+$3710	&	149.94	&	37.17	&	3.166	& 5BZB J1004$+$3752	& 0.440        	&	1.21	\\
IC J1535$+$3807	&	233.79	&	38.11	&	3.073	& 5BZU J1536$+$3742	& 0.679        	&	0.48	\\
IC J1619$+$3841	&	244.86	&	38.68	&	3.734	& 5BZQ J1617$+$3801	& 1.60867$^{B}$	&	0.73	\\
IC J1210$+$3939	&	182.46	&	39.64	&	3.915	& 5BZB J1210$+$3929$^{a,d}$	& 0.61693$^{B}$	&	0.20	\\
IC J1244$+$4037	&	191.07	&	40.62	&	3.644	& 5BZQ J1243$+$4043	& 1.518       	&	0.13	\\
IC J1201$+$4213	&	180.18	&	42.21	&	4.791	& 5BZG J1156$+$4238	& 0.17162$^{B}$	&	0.84	\\
IC J1126$+$4249	&	171.57	&	42.81	&	3.332	& 5BZB J1122$+$4316	& 0.43516$^{B}$	&	0.88	\\
IC J1742$+$4336	&	265.45	&	43.61	&	3.133	& 5BZQ J1740$+$4348	& 2.246        	&	0.26	\\
IC J2249$+$4436	&	342.33	&	44.60	&	3.214	& 5BZB J2247$+$4413$^{a}$	& $>$0.31$^{I}$	&	0.46	\\
IC J1523$+$4559	&	230.74	&	45.98	&	3.490	& 5BZB J1523$+$4606	& --   	        &	0.15	\\
IC J0805$+$5005	&	121.33	&	50.09	&	4.005	& 5BZQ J0808$+$4950$^{a,d}$	& 1.4344$^{B}$	&	0.59	\\
IC J1327$+$5029	&	201.86	&	50.48	&	3.638	& 5BZQ J1327$+$5008	& 1.01191$^{J}$	&	0.33	\\
IC J1644$+$5052	&	250.93	&	50.87	&	3.397	& 5BZU J1647$+$4950$^{a}$	& 0.0475       	&	1.20	\\
IC J0512$+$5701	&	78.12	&	57.02	&	3.671	& 5BZQ J0514$+$5602	&	2.19      	&	1.01	\\
IC J0546$+$5844	&	86.54	&	58.73	&	3.234	& 5BZB J0540$+$5823$^{a}$	& $>$0.1$^{K}$	&	0.81	\\
IC J1855$+$6531	&	283.87	&	65.51	&	3.041	& 5BZB J1848$+$6537$^{a}$	&	0.364    	&	0.74	\\
IC J1938$+$7330	&	294.50	&	73.50	&	3.548	& 5BZQ J1927$+$7358	& 0.302         &	0.85	\\
IC J1356$+$7744	&	208.88	&	77.73	&	3.052	& 5BZQ J1357$+$7643$^{a}$	&	1.585	    &	1.02	\\
\cline{1-7}
\botrule
\end{tabular}}
\begin{tablenotes}
    \begin{minipage}{0.9\textwidth}
    \vspace{0.1cm}
	\small 
	\item
$^{a}$ Objects listed in 4FGL-DR3; $^{b}$ 5BZB J0509$+$0541, a.k.a.  \TXS ; $^{c}$ 5BZB J1427+2348, a.k.a. PKS~1424$+$240; $^{d}$ An excess of neutrino events in the direction of 5BZB J1210$+$3929 and 5BZQ J0808$+$4950 was yet reported in \citet{smith:2021}.; Redshifts are from : 
$^{A}$ \cite{Dunlop:1989}; $^{B}$ SDSS-DR17; $^{C}$ \cite{Ajello_TXS_redshift:2014}; $^{D}$ \cite{LAMOST}; $^{E}$ \cite{Arsioli:2015}; $^{F}$ \citet{Truebenbach:2017}; $^{G}$ \cite{ZBLLAC}; $^{H}$ \citet{Paggi:2014}; $^{I}$ \cite{Shaw:2013}; $^{J}$ \cite{Hewett:2010}; $^{K}$ \cite{Paiano:2020}. 
    \end{minipage}
\end{tablenotes}
\end{threeparttable}
\end{center}
\end{table*}


%
%
\begin{table*}
\begin{center}
\caption{
List of high-galactic latitude ($|b| > 10^\circ$) neutrino hotspots without a candidate 5BZCat association located within the association radius found by this study. 
}\label{tab:unassociated_spots}
\begin{threeparttable}
\resizebox{0.5\textwidth}{!}{ 
\begin{tabular}{@{}lrcc@{}}
\toprule
&   $\alpha_{hs}$[$^{\circ}$]   &   $\delta_{hs}$[$^{\circ}$]   & $L$  \\ 
\cline{1-4}
IC J2040$+$0036	&	309.90	&	$-$0.60	&	3.494	\\
IC J2011$+$0147	&	302.87	&	1.79	&	3.607	\\
IC J1423$+$0232	&	215.86	&	2.54	&	3.061	\\
IC J0039$+$0729	&	9.67	&	7.48	&	4.342	\\
IC J0938$+$0927	&	144.49	&	9.44	&	3.604	\\
IC J0721$+$1125	&	110.21	&	11.42	&	4.039	\\
IC J2127$+$1238	&	321.86	&	12.64	&	3.129	\\
IC J2326$+$1306	&	351.56	&	13.09	&	3.405	\\
IC J0227$+$1524	&	36.74	&	15.40	&	3.602	\\
IC J1621$+$1601	&	245.21	&	16.02	&	3.804	\\
IC J2027$+$1909	&	306.74	&	19.16	&	3.223	\\
IC J0148$+$2045	&	26.89	&	20.74	&	3.116	\\
IC J0811$+$2211	&	122.70	&	22.19	&	3.060	\\
IC J1353$+$2309	&	208.12	&	23.16	&	4.579	\\
IC J1522$+$2319	&	230.45	&	23.32	&	3.212	\\
IC J1353$+$2617	&	208.30	&	26.28	&	3.692	\\
IC J1121$+$2747	&	170.16	&	27.78	&	3.833	\\
IC J0301$+$3144	&	45.18	&	31.74	&	3.138	\\
IC J1331$+$3352	&	202.68	&	33.87	&	4.013	\\
IC J1606$+$3829	&	241.52	&	38.49	&	3.073	\\
IC J1157$+$5225	&	179.33	&	52.42	&	4.221	\\
IC J2129$+$6756	&	322.12	&	67.93	&	4.029	\\
IC J0106$+$7330	&	16.50	&	73.50	&	3.011	\\
IC J1422$+$7616	&	215.40	&	76.26	&	3.720	\\
\cline{1-4}
\botrule
\end{tabular}}
\end{threeparttable}
\end{center}
\end{table*}

%
%
\begin{table*}
\begin{center}
\caption{Neutrino hotspots 
    located close to the the Galactic plane ($|b|\leq 10^{\circ}$). The top (bottom) part of the table lists the northern (southern) hemisphere galactic hotspots, found in \originII\ (\origin). These hotspots are not included in the cross-correlation analysis due to the deficit of blazars near the galactic plane and their possible galactic origin (see Appendix \ref{sec:galplane_hotspots}). The columns report the name, equatorial coordinates (J2000) and the $L$ value of the hotspot. For reference, the right columns report candidate 5BZCat blazar counterpart that are found a-posteriori within the association radius of \originII.
    } 
    \label{tab:galplane_hotspots}
\begin{threeparttable}
\resizebox{0.9\textwidth}{!}{ 
\begin{tabular}{@{}lrcclcc@{}}
\toprule
IceCube hotspots &&&& Potential blazar counterpart&\\
\hhline{=======}
&   $\alpha_{hs}$[$^{\circ}$]   &   $\delta_{hs}$[$^{\circ}$]   & $L$ & 5BZCat & $z$ & Separation[$^{\circ}$] \\ 
\cline{1-7}

IC J0703$+$0103	        &	105.82	&	1.04	&	4.315	&	--		            &	--	    &	--	     \\
IC J1931$+$0138	        &	292.85	&	1.64	&	3.334	&	--		            &	--	    &	--	     \\
IC J0638$+$1039	        &	99.49	&	10.66	&	3.136	&	--		            &	--	    &	--	     \\
IC J1914$+$1524	        &	288.46	&	15.40	&	3.118	&	5BZU J1911$+$1611	&	--      &	0.91	 \\
IC J1857$+$2339	        &	284.24	&	23.64	&	3.092	&	--		            &	--	    &	--	     \\
IC J1950$+$2727	        &	297.42	&	27.45	&	5.275	&	--		            &	--	    &	--	     \\
IC J1958$+$2950	        &	299.53	&	29.83	&	4.187	&	--		            &	--	    &	--	     \\
IC J0530$+$3248	        &	82.44	&	32.80	&	4.049	&	--		            &	--  	&	--	     \\
IC J1930$+$3320	        &	292.50	&	33.33	&	3.533	&	--		            &	--     	&	--	     \\
IC J0433$+$3852$^{a}$	&	68.20	&	38.87	&	3.220   &	--		            &	--	    &	--	     \\
IC J0433$+$4025	        &	68.20	&	40.42	&	3.066	&	--		            &	--	    &	--	     \\
IC J2030$+$4324	        &	307.56	&	43.41	&	3.219	&	--		            &	--	    &	--	     \\
IC J2024$+$4523	        &	306.11	&	45.39	&	3.471	&	--		            &	--	    &	--	     \\
IC J2239$+$4808	        &	339.71	&	48.14	&	3.214	&	--		            &	--	    &	--	     \\
IC J0333$+$5127	        &	53.26	&	51.45	&	3.253	&	--		            &	--	    &	--	     \\
IC J0236$+$6627	        &	39.02	&	66.44	&	3.090	&	5BZQ J0228$+$6721	&	0.523	&	1.15	 \\
\midrule 
IC J0719$-$2240     &	109.78	&	$-$22.67	&	4.081	&	--	&	--	 &	--	 \\
IC J1604$-$5735     &	241.07	&	$-$57.59	&	4.724	&	--	&	--	&	--	 \\
\cline{1-7}
\botrule
\end{tabular}}

\end{threeparttable}
\end{center}
\end{table*}


\bibliography{main}{}
\bibliographystyle{aasjournal}
%
%
\end{document}